\documentclass[useAMS,usenatbib]{mn2e}
\usepackage{graphicx}
\usepackage{graphics,float}                         
\usepackage{epsf}
\usepackage{amsmath}                                
\usepackage{amssymb}
\usepackage{picinpar}                               
\usepackage{relsize,setspace,subfigure}
\usepackage{tabularx}
\usepackage[innercaption]{sidecap}                  
\usepackage[figuresright]{rotating}
\usepackage{url}
\usepackage{aas_macros}
\usepackage{natbib}

\onecolumn

\title[Fisher Information and Wave Interpretations of Universe]{Connection between Fisher Information and Wave Mechanical Interpretations of Universe}
\author[Russell \& Pashaev]{Esra Russell$^{1,2}$\thanks{E-mail: esrarussell@iyte.edu.tr}, Oktay K. Pashaev$^{1}$
\\
$^{1}$Department of Mathematics, Izmir Institute of Technology, 35430 Gulbahcekoyu/Urla, Izmir, Turkey.
\\
$^{2}$Kapteyn Astronomical Institute, University of Groningen, PO
Box 800, 9700 AM Groningen, The Netherlands.}

\begin{document}\date{Accepted .... Received ...; in original form ...}

\pagerange{\pageref{firstpage}--\pageref{lastpage}} \pubyear{2013}

\maketitle

\label{firstpage}
\begin{abstract}
In this study, we model the dark matter and baryon matter distribution in the Cosmic Web by means of
highly nonlinear Schr\"{o}dinger type and reaction diffusion wave mechanical descriptions. The construction of these wave mechanical models of the structure formation is achieved by introducing the Fisher information measure and its comparison with highly nonlinear term called the quantum potential in the wave equations. Strikingly, the comparison of the nonlinear term and the Fisher information measure provides a dynamical distinction between lack of self-organization and self-organization in the dynamical evolution of the cosmic components.
Mathematically equivalent to the standard cosmic fluid equations, these approaches make it possible to follow the
evolution of the matter distribution even into the highly nonlinear regime
by circumventing singularities. In addition, these wave formalisms are extended to two-fluid descriptions of the
coupled dark matter and baryon matter distributions in the linear regime, in the Einstein de Sitter Universe (EdS) to construct toy models of the cosmic components in this relatively simple Universe model. Based on these two different wave mechanical formalisms, here fully analytical results for the dark matter and baryon distributions are provided. Also, numerical realizations of the emerging weblike patterns are presented from the nonlinear dynamics of the baryon component corresponding to soliton-like solutions. These soliton-like solutions might represent a proper description of filamentary structures even in the linear regime.
\end{abstract}
\begin{keywords}
methods: analytical --cosmology: theory, dark matter, large-scale structure of Universe
\end{keywords}
\section{Introduction}
The large scale structure of the Universe is marked by prominent filamentary features embedded within a weblike
network, the Cosmic Web \citep{bond}. Extensive $N$-body simulations are used to model and understand its complex and intricate dynamical structure. The N-body simulations are based on semi-analytical models and the two well known theoretical methods. These methods are classified into two broad classes: The Eulerian and the Zel'dovich approximations.

While the Eulerian approximation provides an accurate description of the gravitational instability in the linear regime, the Zel'dovich approximation is an exact solution of the fluid equations as long as particle trajectories do not cross each other \citep{zeldovich72}. When the trajectories cross, the velocity field becomes multi-valued by causing singularities in the density field. To solve this singularity problem, the adhesion theory is proposed by \cite{KofmanShandarin}. In the adhesion approximation, when shell crossing occurs, the particles are assumed to stick to each other by introducing an artificial viscosity term in the Burger's equation. In the special case, when the viscosity term tends to zero, structures formed in the adhesion model are infinitely thin and the adhesion approximation reduces to the Zel'dovich approximation outside of mass concentration. As is seen, these analytical models are not enough to describe full nonlinear evolution of the structure formation of the Universe. That is why we may need to find a full analytical formalism to understand its complex structure.

As an alternative approach, \cite{Spiegel80} show the correspondence between the fluid structure equations and the nonlinear wave equations. Approximately a decade after this work, \cite{wk93} are the first to apply the Schr\"{o}dinger representation to the problem of the cosmological structure formation for cold dark matter (CDM). \cite{wk93} develop an advanced nonlinear numerical method known as the Schr\"{o}dinger Method (SM) to follow the nonlinear evolution of the dark matter field by introducing an alternative particle mesh code. This new numerical model describes the matter as a Schr\"{o}dinger field obeying the coupled classical Schr\"{o}dinger and Poisson equations. This code is later modified by \cite{dw97}. Another extension of the SM is done by \cite{coles02}. \cite{coles02} suggests that the nonlinear Schr\"{o}dinger equation is a good candidate to model the CDM. However, \cite{coles02} points out that this nonlinear equation presents some difficulties to model the CDM. Later on, \cite{sk03} introduce an elegant nonlinear Schr\"{o}dinger cosmological dark matter perturbation theory in the correspondence limit that gives a formalism equivalent to the collisionless Boltzmann (or Vlasov) equations.

Following up on the work of \cite{coles02}, \cite{colespencer03} demonstrate a wave mechanical approach to treat the singularity problem of the Zel'dovich approximation. This approach is similar to the adhesion approach. \cite{colespencer03} obtains a nonlinear dynamical term which is analogous to the infamous quantum pressure (or quantum potential) and, suggest that this nonlinear term has the same effect as the viscosity term of the adhesion theory. As a result of this, the wave mechanical approach avoids the singularities of the density field. In their study, \cite{colespencer03} also investigate the effect of the quantum pressure term in the gravitational instability. Based on this study of \cite{colespencer03}, \cite{sc06,sc06b} propose a different approach of self gravitating CDM called the free particle approximation. In this approach, quantum pressure and the gravitational potential are neglected. \cite{sc06,sc06b} transform the usual hydrodynamical equations of motion into a linear Schr\"{o}dinger equation. Also they show that the free particle approximation is useful into the mildly nonlinear regime and it has the same result as the adhesion approximation. Another important alternative work on the Schr\"{o}dinger approach in order to interpret the dark matter evolution is done by \cite{jlh2010}. Here, the wave mechanical solutions of the equations of motion for the cosmological homogeneous background evolution of a spherical dark matter overdensity are first obtained, in which the effect of the so called quantum potential is neglected. The reason of ignoring the quantum potential is explained by an assumption that on large scales this nonlinear quantum potential term becomes unimportant. Then \cite{jlh2010} obtain the boundary conditions satisfied by the wave function that is analyzed from the quantum mechanical point of view. Following this, in the concept of quantum mechanics, they find the equations governing the evolution of multiple fluids and then solve them numerically in such a system. Note that \cite{wk93,dw97,colespencer03,sc06,sc06b,jlh2010} use the Madelung transformation of \cite{Madelung27} to obtain Schr\"{o}dinger equation from the cosmological fluid dynamical equations.

It is important to mention that the nonlinear quantum potential term also arises in the different studies that are slightly different due to its derivation. For example, the nonlinear quantum pressure emerges out of exact solutions of the full Wheeler-De Witt equation \citep{blaut98}. This is especially important given the fact that the cosmic wave function describing the quantum state of the Universe satisfies the cosmological kinetic Wheeler-De Witt equation. Apart from this, \cite{lan99} show that the quantum pressure in the causal interpretation of quantum mechanics can be reduced to the Lane-Emden equation that is discussed extensively in the theory of stellar evolution \citep{ch39,cade11,bVg12,b12,caha12,hmo2012}. Moreover, \cite{rn03} suggest a solution to the cosmological problem of the formation and evolution of gravitational structures on many scales by using a gravitational Schr\"{o}dinger equation. Its solutions give probability densities that quantitatively describe precise morphologies in position space and in velocity space \citep{rn03}. Finally the theoretical predictions are successfully checked by a comparison
with observational data, and it is found that matter is self-organized in accordance with the solutions of the gravitational Schr\"{o}dinger equation \citep{rn03}. In addition, \cite{LeeKoh96,Peebles2000,Goodman2000,Arbey2003,boha07} suggest that dark matter haloes may be the self-gravitating Bose-Einstein condensates (BCEs) with short range interactions by a single wave function $\psi(\mathbf{r},t)$. \cite{boha07} also point out that this wave function obeys the Gross-Pitaevskii-Poisson system. Following these studies, \cite{vortex2012} mention that phase-space density of light bosons as dark matter candidates may form BECs. Therefore they show that vortex formation in haloes can be described as a fluid by obeying the cubic nonlinear Schr\"{o}dinger equation (NLSE) which is also known as Gross-Pitaevskii equation. Although this concept of light bosons obeying BEC as dark matter candidates is not well studied yet, there are some other recent results on structure formation studies involving BEC as cold dark matter \citep{guzman2003,fmt08,mf10,harko11,harko11tez}. \cite{wc09} indicate that quantum pressure results from quantum stress and this pressure acts against gravity for the light bosons in the high resolution simulations. In the same study, they confirm that low-mass halos are indeed suppressed by quantum stress even when the small scale fluctuations are abundant in the initial power spectrum. Also, \cite{cavanis12} obtains the quantum potential in the cosmological fluid equations. However by assuming the Thomas-Fermi limit (large mass bosons), \cite{cavanis12} ignores its contribution in the equations.

In addition, all these studies neglect the effect of the nonlinear term the so called quantum pressure (or quantum potential) in different type of Schr\"{o}dinger equations arisen from the astrophysical processes. As is aforementioned, the reason to neglect this is shown as the scale of interest, which is a large scale state, not a microscopic state of quantum mechanics. Also, \cite{colespencer03} and \cite{jlh2010} point out that this nonlinear term is dependent on the amplitude of the wave and the density field slowly varies in large scale structures. On the other hand, from the quantum mechanical point of view, the quantum potential term is not dependent on the density of the field  but only upon its form. As a result of this property, this nonlinear term may show a strong effect on the motion of a particle where the density is very low. This form dependence leads to strong effects even for small particles that are separated by large distances \citep{bh93,bh94,hm00}. Because of these strong effects, in the concept of quantum mechanics, \cite{bh93} and \cite{bh94} suggest that the quantum potential may be interpreted as information potential. The word information stands for an action that brings the order and self organization. In the case of analogy between particles and an $N$-body system of the Universe, the self organization nature of this nonlinear term involves a nonlocal correlation of motion of all the bodies in the collective density field \citep{bh93,bh94,hm00,HileyMaroney2000}. Therefore, $N$-bodies may be controlled by a pool of information that is encoded in the wave function. Note that this pool of information guides the system of bodies rather that affecting them mechanically \citep{hm00,HileyMaroney2000}. Therefore, ignoring of emergence of the quantum potential as a self organization may limit our way to to obtain a full understanding of evolution of the large scale structure of the Universe, since it indicates an important role in the large scale evolution of the Universe rather than being an unimportant microscopic parameter. Consequently, it seems that the main reason of skipping this term from the wave equations is caused by mathematical tractability rather than its physical interpretation.

Separately from the cosmological perspective, \cite{parwanipashaev} relate the quantum potential/pressure term with the positive signed Fisher Information Measure of information theory. They also show that depending on the choice of the enthalpy function which is related with the equation of state, one may obtain the cubic NLSE or other modified NLSEs for barotropic compressible fluids. On the other hand, \cite{2009arXiv0901.3742F} proves that a negative sign of FI leads to maximizing the FI which is equivalent to the Shannon measure in the concept of biological systems. From the dynamical point of view, \cite{Cabezas20023} relates the Fisher information with dynamical behavior of a system based on the sign of Fisher information. According to this, positive FI indicates loss-of self organization while negative one demonstrates strong self-organization in a dynamical system. Apart from this, \cite{1964mhdp.book.....M} indicate that the linearization of any fluid equation can be done by a choice of density distribution in the form of a real or an imaginary exponential transformations. These transformations lead to the two different dynamical wave interpretations of the same fluid, which are the Schr\"{o}dinger type and reaction diffusion systems.

In this paper, progressing by previous studies, we derive fully analytical wave mechanical approaches of the dark matter and baryon field components of the Universe in a two-fluid formalism in the linear regime. To do this, first we introduce the cosmological Newtonian fluid dynamical equations in comoving coordinates in order to give the evolution of the Universe in terms of expanding background. Following this, comoving matter fluid equations are scaled by using velocity potential and scaled density parameters. Then we show that matter fluid obeys a Schr\"{o}dinger type nonlinear wave form which has contributions from baryon and dark matter fields and includes the infamous quantum pressure term by applying the Madelung transformation \citep{Madelung27}. After introducing the Fisher information measure in the Lagrangian functional of the matter component that obey the Schr\"{o}dinger type equation, it is shown that two different wave forms arise for the two-fluid formalisms depending on sign given by the relation of the Fisher measure of the system and the guiding force (or quantum potential) of the same system. As a result, it is indicated that the Newtonian fluid dynamical equations of the incompressible cosmological two component fluid can be transferred into nonlinear Schr\"{o}dinger equations as well as Reaction diffusion/heat equations based on the study of \cite{1964mhdp.book.....M}. Then we provide full solutions of these equations with the nonlinear potential term in the EdS Universe in the linear regime. Moreover, it is also showed that the solution of these formalisms provides the Zel'dovich approach for the Schr\"{o}dinger wave description of the dark matter component in the linear regime. On the other hand, the baryon component presents soliton-like perturbative solutions from the Reaction Diffusion type wave description.

\section{Modified Fluid Dynamical Equations}
To obtain the nonlinear Schr\"{o}dinger type wave equations to model the dark matter and baryon components, we take our first starting point as the Newtonian equations for a matter fluid in terms of comoving coordinates in an expanding Universe. Then the cosmological fluid equations in comoving coordinates can be written as the continuity,

\begin{eqnarray}
\frac{\partial \delta}{\partial t}+\frac{1}{a}\mathbf{\nabla_{x}}\left[\left(1+\delta\right)\mathbf{v}\right]=0,
\label{cont1}\end{eqnarray}
the Euler,
\begin{eqnarray}
\frac{\partial \mathbf{v}}{\partial t}+H \mathbf{v}+\frac{1}{a}\mathbf{v}\mathbf{\nabla_{x}.v}=-\frac{1}{a}\mathbf{\nabla_{x}}\phi-\frac{1}{a}\frac{1}{\rho_{u}\left(1+\delta\right)}\nabla_{x}p,
\label{euler1}\end{eqnarray}
and the Poisson equation,
\begin{eqnarray}
\mathbf{\nabla_{x}}^2\phi=4\pi G a^2\rho_{u}\left(1+\delta\right),
\label{poisson1}\end{eqnarray}
where $x$ is the comoving coordinate expanding with the scale factor $a(t)$, the Hubble parameter $H = H(t)$ is defined by $H= \dot{a}/{a}$ where the dot denotes a derivative with respect to time $t$. The peculiar velocity field
$\mathbf{v} = \mathbf{v}(\mathbf{x}, t)$ is given by $\mathbf{v} = a \dot{\mathbf{x}}$ and $\phi = \phi(\mathbf{x}, t)$ is the peculiar Newtonian
gravitational potential. The density contrast $\delta= \delta(x, t)$ is $\delta+1 = \rho/\rho_{u}$ where $\rho = \rho(\mathbf{x}, t)$ is the matter density field, while $\rho_{u} = \rho_{u}(t)$ is the density contribution in the homogeneous background. The pressure function is demonstrated by $p(\mathbf{x},t)$. The relation between pressure and the density $\rho(\mathbf{x}, t)$ is given by the equation of state,

\begin{eqnarray}
p=\omega \rho,
\label{eos}\end{eqnarray}
where $\omega$ is the dimensionless adiabatic parameter and characterizes the constitute of the medium. The cosmological perfect fluid shows a barotropic process in which the pressure is only a function of density $p=p(\rho)$. The total energy of the perfect fluid that fills the Universe is defined by the enthalpy function which can be written in terms of enthalpy potential $V_{enth}(\rho)$ and is given by \cite{parwanipashaev} as,

\begin{eqnarray}
\epsilon(\rho)=\int^{\rho}_{\rho_{0}}\frac{d p}{\rho},\phantom{a} \epsilon(\rho)=\frac{d V_{enth}(\rho)}{d\rho}.
\label{enthalphy}\end{eqnarray}

\section{Einstein de Sitter Universe}
The Einstein de Sitter (EdS) Universe is a matter dominated Friedmann model with a flat geometry since its critical density is equal to unity $\Omega=1$. Due to $\Omega=1$, the EdS Universe expands forever. In this universe model, the expansion factor $a(t)$ is proportional to,

\begin{eqnarray}
a(t)\propto t^{2/3}\propto \mathcal{D}(t),
\end{eqnarray}
where $\mathcal{D}(t)$ is the linear density growth factor and it indicates the structure formation of the universe.

\section{Scaling the Fluid Dynamical Equations}\label{sec:ScalingFluidEqns}
In addition to these basic definitions and mathematical descriptions, it is sensible to describe the evolution in terms of the expansion factor $a(t)$ or, even more convenient and appropriate, in terms of the linear density growth factor $\mathcal{D}(t)$ \citep{sc06} in order to describe the evolution against an expanding background. Hence, the linear growth factor follows from solving the second order ordinary differential equation \citep{peebles80},

\begin{eqnarray}
\ddot{\mathcal{D}}+2H\dot{\mathcal{D}}-\frac{3}{2}\Omega H^{2} \mathcal{D}=0,
\label{growth}
\end{eqnarray}
\noindent
where the dot presents derivative with respect to $t$ and, at some initial time $t_{i}$, $\mathcal{D}_{i} = \mathcal{D}(t_{i}) = 1$ and the critical density is $\Omega=1$ in the EdS Universe. With respect to the time
variable $\mathcal{D}(t)$, we define a scaled peculiar velocity ${\mathbf{v^{\prime}}}$,

\begin{eqnarray}
{\mathbf{v^{\prime}}}\equiv\frac{dx}{d \mathcal{D}}=\frac{\mathbf{v}}{a\dot{\mathcal{D}}}.
\label{newv1}\end{eqnarray}
\noindent
We also introduce the comoving velocity potential $\phi_{v}$ for the velocity ${\mathbf{v^{\prime}}}$,
\begin{eqnarray}
{\mathbf{v^{\prime}}}\equiv\nabla_{x}\phi_{v},
\label{newv2}\end{eqnarray}
and the scaled density $\chi$ as follows,
\begin{eqnarray}
\chi(x,t)\equiv\delta+1= \rho/\rho_{u}.
\end{eqnarray}
As a result, the velocity flow characterized by the velocity potential into the Euler (\ref{euler1}) and continuity (\ref{cont1}) equations are scaled. In addition, the scaled Euler equation is integrated once in terms of comoving coordinates in order to obtain the Bernouille equation. Hence the scaled continuity and the scaled Bernoullie equations become,

\begin{eqnarray}
\frac{\partial \chi}{\partial \mathcal{D}}+\nabla_{x}\left(\chi\nabla_{x}\phi_{v}\right)=0,
\label{cont2}
\end{eqnarray}

\begin{eqnarray}
\frac{\partial \phi_{v}}{\partial
\mathcal{D}}+\frac{1}{2}\left(\nabla_{x}\phi_{v}\right)^2=-V_{eff}-A^{2}(\mathcal{D})\epsilon(\chi),
\label{bernoullimod}
\end{eqnarray}
where $\epsilon(\chi)$ is the scaled enthalpy. The time dependent function $A$ is defined as $A(\mathcal{D})\equiv 1/a\dot{\mathcal{D}}$ in which scale factor $a$ is equivalent to the linear growth factor $\mathcal{D}$ in the EdS Universe. Then the dispersion term $A$ becomes dependent on the growth factor $\mathcal{D}$ only. Apart from this, the effective potential $V_{eff}$ includes contributions from the matter velocity potential $\phi_{v}$ and the gravity potential $\phi$,
\begin{eqnarray}\label{effectivepot}
V_{eff}&=&\frac{3}{2f^2 \mathcal{D}}\Omega\phi_{v}+\frac{1}{a^2\dot{\mathcal{D}}^2}\phi=\frac{3\Omega}{2f^2 \mathcal{D}}\left(\phi_{v}+\theta_{g}\right).
\end{eqnarray}
\noindent
Here, the linear velocity growth factor $f(\Omega)$, also known as the Peebles factor \cite{peebles80} is given as,

\begin{eqnarray}
f =\frac{a\dot{D}}{\dot{a}D}=\frac{1}{H}\frac{\dot{D}}{D},
\end{eqnarray}
and the scaled gravity potential $\theta_{g}$ defined as,

\begin{eqnarray}
\theta_{g}=\frac{2\phi}{3\Omega a^2 \mathcal{D} H^2}.
\end{eqnarray}
In the following section, we introduce the two-fluid wave mechanical approach and discuss the emergence of two different wave formalisms for each cosmic component.

\section{Effective Potential in the EdS Universe at the linear regime $\delta\ll 1$}\label{sec:effectivepotinEDS}
As is aforementioned in the section \ref{sec:ScalingFluidEqns} in its definition (\ref{effectivepot}), the effective potential have the total contribution form of the peculiar gravitational function,

\begin{eqnarray}
V_{eff}=V_{m}=\frac{3\Omega\phi_{v,m}}{2f^2 \mathcal{D}}+\frac{\phi_{m}}{a^{2}\dot{\mathcal{D}}^{2}},
\nonumber
\end{eqnarray}
\noindent
where the effective potential $V_{m}$ includes contributions from the dark matter velocity potential $\phi_{v,dm}$ and the matter gravity potential $\phi_{m}$. \cite{jones99} presents an analytical model for nonlinear clustering of
the baryon material in a Universe where the gravitational field is dominated by dark matter and this baryon
matter flow is dissipative and is driven by the dark matter potential. Therefore, here it is assumed that the potential function is dominated by the dark matter component as dark matter creates ever deeper potential wells in which
baryon matter will fall. This means that in the total peculiar gravitational potential function $\phi_{m}=\phi_{b}+\phi_{dm}$, the baryon component has a very small contribution $\phi_{b}<<\phi_{dm}$ and we can say that the total gravitational field is dominated by the dark matter component $\phi_{m}\approx\phi_{dm}$. As the baryon component is driven by the dark matter gravitational field, we can assume that the velocity field of the baryon matter follows the dark matter velocity field. This means that the velocity potentials of the dark and baryon matter components are approximately equal $\phi_{b,dm}\approx\phi_{v,dm}$. Under these assumptions the effective potentials
become,

\begin{eqnarray}
V_{b}\approx V_{dm}=\frac{3\Omega\phi_{v,dm}}{2f^2 \mathcal{D}}+\frac{\phi_{dm}}{a^2\dot{\mathcal{D}}^2}.
\end{eqnarray}
\noindent
At this point, it is interesting to note that in the linear regime the effective potential $V_{dm}$ is equal to zero. We may easily infer this from the
direct linear relation between peculiar velocity $v_{dm}$ and the peculiar gravity $g$ in the linear regime \citep{peebles80},

\begin{eqnarray}
v_{dm}=\frac{2f}{3\Omega H}g=-\frac{2f}{3\Omega H a}\nabla_{x}\phi_{dm}.
\end{eqnarray}
\noindent
Using this relation in the expression for the dark matter velocity potential $\phi_{v,dm}$, we obtain,

\begin{eqnarray}
\nabla_{x}\phi_{v,dm}={v_{dm}^{\prime}}=\frac{v_{dm}}{a\dot{\mathcal{D}}}=-\frac{2f}{3\Omega H a^2 \mathcal{D}}\nabla_{x}\phi_{dm},
\end{eqnarray}
\noindent
from which we find the linear regime relation between dark matter velocity potential $\phi_{v,dm}$ and the peculiar potential $\phi_{dm}$,

\begin{eqnarray}
\phi_{v,dm}=-\frac{2f}{3\Omega H a^2\mathcal{D}}\phi_{dm}.
\label{velpotdm}
\end{eqnarray}
\noindent
If we rearrange the effective potential by using equation (\ref{velpotdm}),

\begin{eqnarray}
V_{dm}=\frac{3\Omega\phi_{v,dm}}{2f^2 \mathcal{D}}\left(\phi_{v,dm}+\frac{2f}{3\Omega H a^2 \mathcal{D}}\phi_{dm}\right).
\label{dmeffpot}\end{eqnarray}
\noindent
The immediate conclusion is that the effective dark matter potential $V_{dm}$,

\begin{eqnarray}
V_{dm}=0.
\end{eqnarray}
\noindent
In fact, the conclusion of $V_{eff} = 0$ stretches out much further into the quasi-linear regime, for as long as the Zel'dovich formalism still describes the
motion of matter elements in the Universe.
Due to the vanishing effective potential in the EdS universe and, the linear density perturbations satisfying $\delta\ll 1$, the scaled Bernouille equation (\ref{bernoullimod}) turn into a relatively simple form.

\section{Modeling Dark Matter and Baryon Components}
Using the inverse of the Madelung transformation proposed by \cite{Madelung27}, we show that the dark and baryon matter components can be presented as complex scalar fields. The Madelung transformation is given as follows,

\begin{eqnarray}
\psi(\mathbf{x},\mathcal{D})=\sqrt{\chi}e^{i{\phi_{v}/\nu}},
\label{madelung}
\end{eqnarray}
in which the wave function $\psi(x,\mathcal{D})$ is a complex quantity and $\nu$ is the adjustable parameter,
and has the same dimension as velocity potential $\phi_{v}$. Here it is important to point out that in the original work of \cite{Madelung27} the main purpose was to model quantum fluid. That is why he chose the adjustable parameter as the planck constant $\hbar$. As is seen in the Madelung wave form (\ref{madelung}), we do not adopt the quantum scales due to our interest of scale which is the large scale structure of the Universe. As a result of this, the wave forms that will be derived here are not related with quantum scales.

The density function $\chi$ in (\ref{madelung}) satisfies the relation $\chi=\psi\psi^{*}=|\psi|^2$. Hence the modified NLSE and the Poisson system of equations in which matter density obeys in the Madelung form are given as,
\begin{subequations}
\begin{eqnarray}
i\nu\frac{\partial \psi}{\partial \mathcal{D}}+\frac{\nu^{2}}{2}\mathbf{\nabla_{x}}^2\psi-A^{2}(\mathcal{D})\epsilon(|\psi|^2)\psi&=&
\mathcal{P}\psi,\label{nlS}
\\
\mathbf{\nabla_{x}}^{2} \phi(\mathbf{x},\mathcal{D})&=&4\pi G a^{2}\rho_{u}\left|\psi\right|^{2}.
\end{eqnarray}
\end{subequations}
Here the nonlinear term $\mathcal{P}$ on the right hand side of equation (\ref{nlS}) has the analogy with the infamous quantum potential/pressure derived by \cite{b52,b52II} and its form is given as,

\begin{eqnarray}
\mathcal{P}\equiv \frac{\nu^2}{2}\frac{\mathbf{\nabla_{x}}^{2}|\psi|}{|\psi|}.
\end{eqnarray}
In this study, we focus on the information potential interpretation of this nonlinear term as is suggested by \cite{bh93} and \cite{bh94}. This interpretation allows us to discuss this nonlinear dynamical term from the perspective of the large scale structure of the Universe as a parameter that brings order and self-organization to the system. Before introducing the Fisher information measure, here we obtain the Lagrangian of the general nonlinear Schr\"{o}dinger type equation (\ref{nlS}),

\begin{eqnarray}
L=i{\nu}\left(\psi^{*}\frac{\partial \psi}{\partial \mathcal{D}}-\psi\frac{\partial \psi^{*}}{\partial \mathcal{D}}\right)+\frac{\nu^{2}}{2}\mathbf{\nabla_{x}}\psi^{*}\mathbf{\nabla_{x}}\psi+V_{enth}(\psi^{*}\psi)
-\frac{\nu^{2}}{8}\frac{\left({\nabla_{x}}|\psi|^2\right)^2}{|\psi|^2}
\label{lag1}
\end{eqnarray}
As is seen, the Lagrangian varies, with respect to scaled density $\chi$ and the velocity potential $\phi_{v}$. Hence, the fluid dynamical equations (\ref{cont2}) and (\ref{bernoullimod}) appear be varying this Lagrangian functional. Following up on \cite{parwanipashaev}, the Lagrangian functional in equation (\ref{lag1}) of the classical equation of motion can be modified by introducing the Fisher measure $I_{F}$ of information theory,

\begin{eqnarray}
\tilde{L}=L+\frac{\lambda^{2}}{8} I_{F},
\label{lag2a}
\end{eqnarray}
where $I_{F}$ is the Fisher information measure and is defined as,

\begin{eqnarray}
I_{F}=\int d\mathcal{D} d\mathbf{x} \chi \left(\mathbf{\nabla_{x}}\log \chi \right)^2 =4\int  d\mathcal{D} d\mathbf{x} \left(\mathbf{\nabla_{x}}\sqrt{\rho}\right)^2.
\label{fishermeasure}
\end{eqnarray}
The main reason of adding the Fisher Information to the Lagrangian is to compare the intrinsic amount of information that is contained by matter component of the Universe and the information that we observe or gain by following the description and interpretation of Fisher information based on previous studies \citep{Frieden2004,Frieden2002,Cabezas20023,2009arXiv0901.3742F}. Here Fisher's information measure $I_{F}$ reflects the amount of information of the observer and it depends on the density $\rho$. The density function $\rho$ appears in this context because there is uncertainty in our knowledge of observing the matter distribution of the Universe as a whole. Therefore one may adopt the principle of maximum uncertainty to constrain the probability distribution $\rho$ characterizing the ensemble: we would like to be as unbiased as possible in its choice, consistent with our lack of information. Hence we modify the Lagrangian by adding the Fisher measure, then the new Lagrangian becomes,
\begin{eqnarray}
L=i{\nu}\left(\psi^{*}\frac{\partial \psi}{\partial \mathcal{D}}-\psi\frac{\partial \psi^{*}}{\partial \mathcal{D}}\right)+\frac{\nu^{2}}{2}\nabla_{x}\psi^{*}\nabla_{x}\psi
+V_{enth}(\psi^{*}\psi)-\frac{\left(\nu^{2}-\lambda^2\right)}{8}
\frac{\left({\nabla_{x}}|\psi|^2\right)^2}{|\psi|^2}
\label{lag2}
\end{eqnarray}
Here the constraint is implemented in the Lagrangian density $\lambda^2/8$ \citep{parwanipashaev} which minimizes the classical action and is called the Lagrange multiplier. Taking into account the modified Lagrangian (\ref{lag2}), the nonlinear Schr\"{o}dinger type equation changes its form as,
\begin{eqnarray}
\label{nlselast}
i\nu\frac{\partial \psi}{\partial \mathcal{D}}+\frac{\nu^{2}}{2}\mathbf{\nabla_{x}}^2\psi-A^{2}(\mathcal{D})\epsilon(|\psi|^2)\psi =\frac{\nu^2-\lambda^2}{2}\frac{\mathbf{\nabla_{x}}^{2}|\psi|}{|\psi|}\psi.
\end{eqnarray}
Here $\nu$ and $\lambda$ are constant parameters. Note that the wave representation of the matter Madelung representation has contributions from dark matter $\Psi_{dm}$ and baryon $\Psi_{b}$ components. As a result, the matter wave function $\Psi_{m}$ becomes,
\begin{eqnarray}
\Psi_{m}\equiv\Psi_{dm}+\Psi_{b}.
\end{eqnarray}
It is crucial to mention that, the equation of state of cosmological fluid (\ref{eos}) is characterized by the constitute of the Universe via the adiabatic parameter $\omega$. This leads to the different enthalpy parameters for different dominant constitutes of the Universe as a result different dynamical behaviors,

\begin{itemize}
\item Dark Matter: $\omega=0$,\phantom{a}$\epsilon\left(\chi_{dm}\right)=0$,
\item Baryon Component: $\omega > 0$,\phantom{a}$\epsilon\left(\chi_{b}\right)=\omega\ln\chi_{b}=\omega\ln\left|\psi_{b}\right|^2$.
\end{itemize}
Therefore, depending on what component dominates the evolution of the Universe, the nonlinear equation can be split into the two dynamical systems,

\begin{eqnarray}
\label{nlselastdm}
i\nu\frac{\partial \psi_{dm}}{\partial \mathcal{D}}+\frac{\nu^{2}}{2}\mathbf{\nabla_{x}}^2\psi_{dm} =\frac{\nu^2-\lambda^2}{2}\frac{\mathbf{\nabla_{x}}^{2}|\psi_{dm}|}{|\psi_{dm}|}\psi_{dm}.
\end{eqnarray}

\begin{eqnarray}
\label{nlselastb}
i\nu\frac{\partial \psi_{b}}{\partial \mathcal{D}}+\frac{\nu^{2}}{2}\mathbf{\nabla_{x}}^2\psi-A^{2}(\mathcal{D})\ln\left|\psi_{b}\right|^2\psi_{b} =\frac{\nu^2-\lambda^2}{2}\frac{\mathbf{\nabla_{x}}^{2}|\psi_{b}|}{|\psi_{b}|}\psi_{b}.
\end{eqnarray}

As is seen in equation (\ref{nlselast}), the relation between these parameters allows us to relate the evolution of the system with the information theory. Note that the scaled density $\chi$ is a positive definite and if it is uniform, it tells us that the underlying particles of the fluid could be anywhere: we have no information (minimum information) \citep{parwanipashaev}. If the density is peaked somewhere, we know
that a fluid particle is more likely to be there; that is we have gained information. Thus we require that our scalar
information has the property that it is a positive definite $\lambda > 0$. On the other hand, \cite{2009arXiv0901.3742F} point out that maximizing the Fisher measure gives a negative sign and this leads to derive the Shannon measure. That is why, in the NLS equation (\ref{nlselast}) the relation between parameters $\nu$ and Langrange multiplier $\lambda$ has a crucial role that affects the dynamical behavior of the components. Following \cite{Cabezas20023}, here we make a distinction between two different dynamical behaviors depending on the relation between the system itself and its information which is given by $\nu^2-\lambda^2$ in Lagrangian equation (\ref{lag2}). According to this, \cite{Cabezas20023} describe two main dynamics based on the Fisher information,
\begin{enumerate}
\item Decreasing Fisher information $\lambda < \nu$: This condition leads the system to a loss of
self-organization and this type of dynamical system obeys a Schr\"{o}dinger type wave form.
\item Increasing Fisher information $\lambda > \nu$: This indicates that the system self-organizes itself. This kind of dynamical behavior obeys Heat/Reaction diffusion type of coupled system of equations.
\end{enumerate}
In the following subsections, these two different dynamical behaviors of the two-fluid formalism are derived and analytical derivations are given.

\subsection{$\lambda < \nu$ Schr\"{o}dinger type wave mechanics: loss of self-organization}
If the information that we observe from the system decreases, then the Fisher information becomes smaller than the self-organization of the system. In this case, $\lambda < \nu$, we obtain the Poisson- NLS type coupled equations for dark matter dark matter, which is,

\begin{subequations}
\label{qqFP}
\begin{eqnarray}
i\frac{\partial \Psi_{dm}}{\partial \tilde{\mathcal{D}}}+\frac{1}{2}\mathbf{\nabla_{x}}^2\Psi_{dm}&=&0,\\
\mathbf{\nabla_{x}}^{2} \phi(\mathbf{x},\tilde{\mathcal{D}})&=&4\pi G a^2\rho_{u}\left|\Psi_{dm}\right|^{2},
\end{eqnarray}
\end{subequations}
and baryon component,

\begin{subequations}
\label{Schb}
\begin{eqnarray}
i\frac{\partial \Psi_{b}}{\partial \tilde{\mathcal{D}}}+\frac{1}{2}\mathbf{\nabla_{x}}^2\Psi_{b}
-\omega\frac{A^{2}(\tilde{\mathcal{D}})}{\nu^{\prime}}\ln|\Psi_{b}|^2\Psi_{b}&=&0,\\
\mathbf{\nabla_{x}}^{2} \phi_{b}(\mathbf{x},\tilde{\mathcal{D}})&=&4\pi G a^2\rho_{u}\left|\Psi_{b}\right|^{2}.
\end{eqnarray}
\end{subequations}

where $\nu^{\prime}$ is defined as,
\begin{eqnarray}
\nu^{\prime}&\equiv&{\nu^2}\left(1-\frac{\lambda^2}{\nu^2}\right).
\end{eqnarray}
and Madelung wave transforms are,
\begin{eqnarray}
\Psi_{dm}=\sqrt{\chi_{dm}} e^{i{\tilde{\phi}_{v}}},\phantom{a}\Psi_{b}=\sqrt{\chi_{b}} e^{i{\tilde{\phi}_{v}}}
\end{eqnarray}
in which $\tilde{\phi}_{v}=\frac{\phi_{v}}{\sqrt{\nu^{\prime}}}$, $\tilde{\mathcal{D}}= \mathcal{D}\sqrt{\nu^{\prime}}$ and $A^{2}(\tilde{\mathcal{D}})$ becomes,

\begin{eqnarray}
A^{2}(\tilde{\mathcal{D}})=\frac{\nu^{\prime}}{\tilde{\mathcal{D}}\tilde{\dot{\mathcal{D}}}}.
\end{eqnarray}

\subsection{$\lambda > \nu$ Reaction Diffusion/Heat type Dynamics: self-organization}
On the other hand, if the information that we obtain from the dynamical system becomes higher than the system's internal self-organization ($\lambda > \nu$), then instead of a complex dynamical form, we get the time reversal pair of reaction diffusion equations of the cosmological fluid. Here we use the special transformation by \cite{1964mhdp.book.....M} and \cite{leepashaev98},

\begin{eqnarray}
Q_{dm}^{+}(\mathbf{x},\tilde{{\tilde{\mathcal{D}}}})\equiv\sqrt{\chi_{dm}}e^{{{\grave{\phi}}}_{v}},\phantom{a} -Q_{dm}^{-}(\mathbf{x},\tilde{\tilde{\mathcal{D}}})\equiv\sqrt{\chi_{dm}}e^{-{{\grave{\phi}}}_{v}},
\end{eqnarray}

\begin{eqnarray}
Q_{b}^{+}(\mathbf{x},\tilde{{\tilde{\mathcal{D}}}})\equiv\sqrt{\chi_{b}}e^{{{\grave{\phi}}}_{v}},\phantom{a} -Q_{b}^{-}(\mathbf{x},\tilde{\tilde{\mathcal{D}}})\equiv\sqrt{\chi_{b}}e^{-{{\grave{\phi}}}_{v}},
\end{eqnarray}
in which ${\grave{\phi}}_{v}=\frac{{\phi}_{v}}{\sqrt{\nu^{\prime\prime}}}$ and $\tilde{\tilde{\mathcal{D}}}= {\mathcal{D}}\sqrt{\nu^{\prime\prime}}$ in which ${\nu^{\prime\prime}}$ is defined as,
\begin{eqnarray}
{\nu^{\prime\prime}}\equiv\nu^2\left(\frac{\lambda^2}{\nu^2}-1\right).
\end{eqnarray}
These real functions and the wave function of the NLS equation satisfy the following relations,

\begin{eqnarray}
-Q^{+}Q^{-}=\chi=\Psi \Psi^{*},\phantom{a}\frac{Q^{+}}{-Q^{-}}=\left(\frac{\Psi}{\Psi^{*}}\right)^{i},
\label{relationqchi}
\end{eqnarray}
Similar to the loss of self-organization case, the contributions of dark matter and baryon contributions in the matter density distributions due to the superposition of the components,

\begin{eqnarray}
Q^{\pm}_{m}=Q^{\pm}_{dm}+Q^{\pm}_{b}.
\end{eqnarray}
\noindent
and considering the different enthalpy parameters for different dominant constitutes of the Universe, are taken into account. Therefore, the reaction diffusion system and the Poisson equation can be split into two different dynamical forms, for the dark matter,

\begin{subequations}
\label{reactionsysdma}
\begin{eqnarray}
\frac{\partial Q^{+}_{dm}}{\partial \tilde{\tilde{\mathcal{D}}}}+\frac{1}{2}\nabla_{\mathbf{x}}^2Q^{+}_{dm}
&=&0,\\
-\frac{\partial Q^{-}_{dm}}{\partial \tilde{\tilde{\mathcal{D}}}}+\frac{1}{2}\nabla_{\mathbf{x}}^2Q^{-}_{dm}&=&0,
\label{reactiondiffeqn2dm}\\
\mathbf{\nabla_{x}}^{2} \phi_{dm}(\mathbf{x},\tilde{\tilde{\mathcal{D}}})&=&4\pi G a^{2}\rho_{u} (-Q^{+}_{dm}Q^{-}_{dm}),
\label{reactiondiffeqngvp2dm}
\end{eqnarray}
\end{subequations}
and the baryon components,

\begin{subequations}
\label{reactionsysb}
\begin{eqnarray}
\frac{\partial Q^{+}_{b}}{\partial \tilde{\tilde{\mathcal{D}}}}+\frac{1}{2}\nabla_{\mathbf{x}}^2Q^{+}_{b}
-\frac{\omega}{\nu^{\prime\prime}}{A^{2}(\tilde{\tilde{{\mathcal{D}}}})}{\ln|-Q^{+}_{b}Q^{-}_{b}|}Q^{+}_{b}&=&0,\\
-\frac{\partial Q^{-}_{b}}{\partial \tilde{\tilde{\mathcal{D}}}}+\frac{1}{2}\nabla_{\mathbf{x}}^2Q^{-}_{b}
-\frac{\omega}{\nu^{\prime\prime}}{A^{2}(\tilde{\tilde{{\mathcal{D}}}})}{\ln|-Q^{+}_{b}Q^{-}_{b}|}Q^{-}_{b}&=&0,
\label{reactiondiffeqn2b}\\
\mathbf{\nabla_{x}}^{2} \phi_{b}(\mathbf{x},\tilde{\tilde{\mathcal{D}}})&=&4\pi G a^{2}\rho_{u} (-Q^{+}_{b}Q^{-}_{b}).
\label{reactiondiffeqn1b}
\end{eqnarray}
\end{subequations}
in which ${A(\tilde{\tilde{{\mathcal{D}}}})}$ is defined as,

\begin{eqnarray}
{A(\tilde{\tilde{{\mathcal{D}}}})}=\frac{\nu^{\prime\prime}}{\tilde{\tilde{{\mathcal{D}}}}\tilde{\tilde{{\dot{\mathcal{D}}}}}}.
\end{eqnarray}
so fluid equations (\ref{cont1}), (\ref{euler1}) or (\ref{bernoullimod}) can also be written as the reaction diffusion systems which are the analog of (\ref{nlS}).

Here we represent the decoupling reaction diffusion systems of the two different cosmic components. Above the reaction diffusion equations with negative signs (\ref{reactiondiffeqn2dm}) and (\ref{reactiondiffeqn2b}) are the time reversible of the positive signed reaction diffusion equations. They are crucial for the existence of Hamiltonian structure and the integrable system \citep{leepashaev98,pashlee02}. Note that the reaction diffusion systems are scaled by following up on \cite{leepashaev98,pashlee02} and in this way the contribution of the Fisher information is hidden in the equations via transformations.

\section{Example A: Dark Matter Dynamics in EdS Universe at $\delta\ll 1$ and $\lambda < \nu$}
\begin{figure}
\centering
\begin{tabular}{cc}
\includegraphics[width=0.4\textwidth]{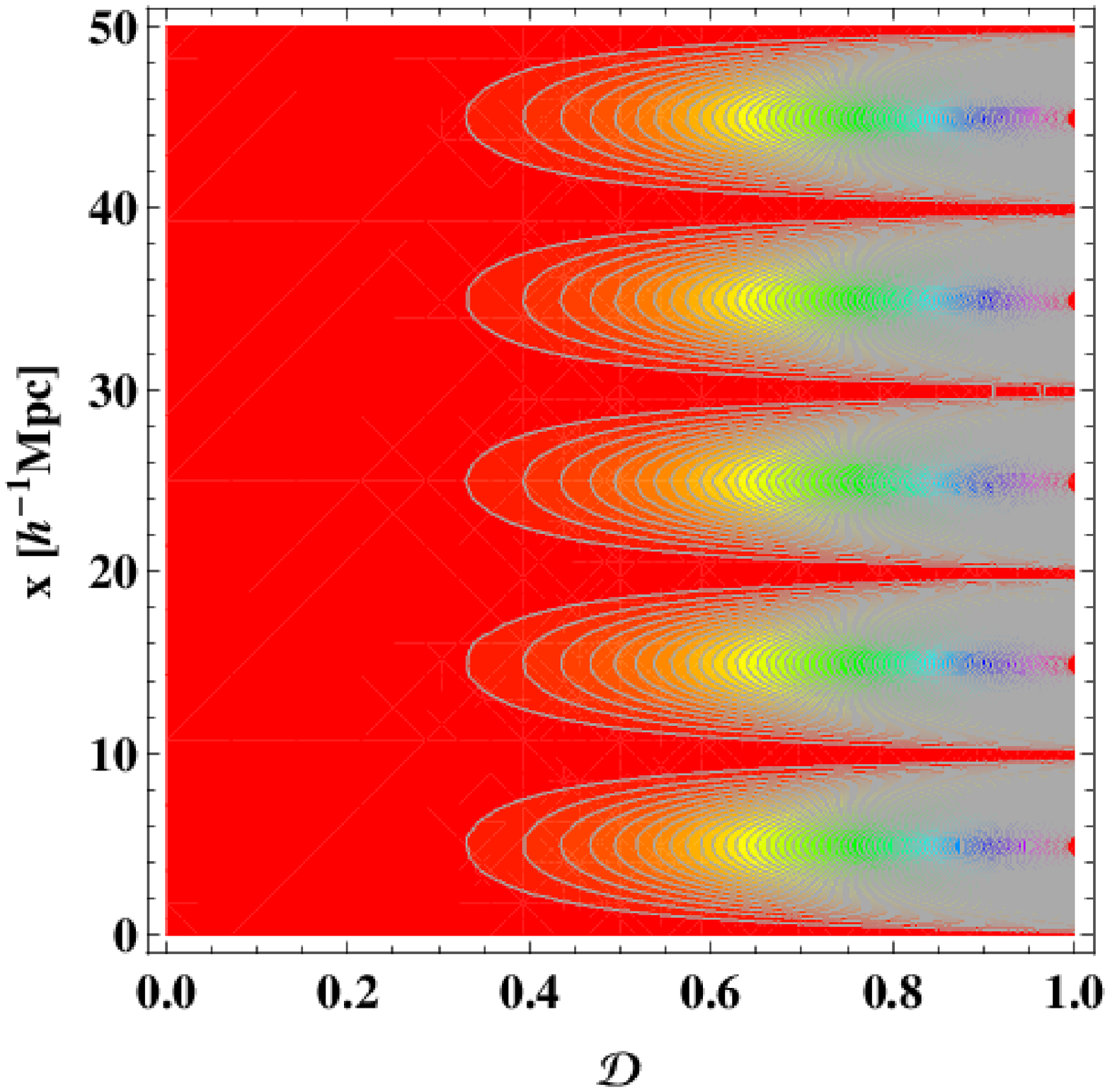}
\includegraphics[width=0.4\textwidth]{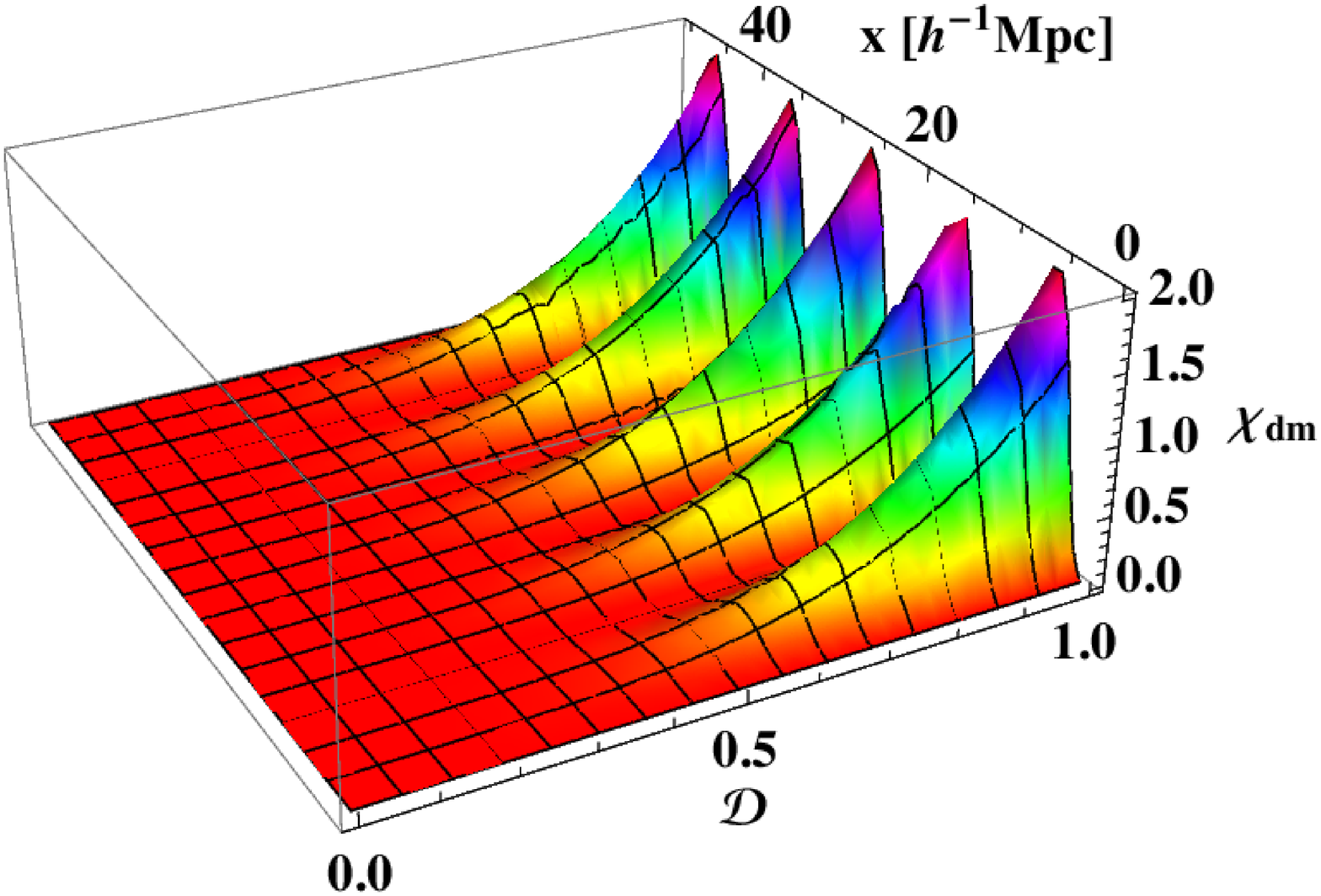}
\end{tabular}
\caption{Contour (left) and three-dimensional (right) plots of the evolution of the scaled dark matter density
function $|\Psi(x, y, z, \mathcal{D})|^2 = \chi_{dm}$ based on equation (\ref{darkmatterdensity}) in terms of
the growth factor $\mathcal{D}$ by choosing parameters as $d_{1}=50$, $d_{2}=50$, $d_{3}=50$, $n_{1}=5$,
$n_{2}=5$ and $n_{3}=1$.}
\label{fig:darkmatter}
\end{figure}

Here, we provide an analytical solution in the case of loss of self-organization $\lambda < \nu$ to
model dark matter as an example. In this case, the dark matter evolution can be modeled by the Schr\"{o}dinger
type wave mechanical approach. In the EdS Universe in terms of the linear regime, the nonlinear Schr\"{o}dinger type wave equation of the dark matter component is reduced to the free particle Schr\"{o}dinger equation as in the system (\ref{qqFP}),

\begin{eqnarray}
i\frac{\partial \Psi_{dm}}{\partial \tilde{\mathcal{D}}}+\frac{1}{2}\mathbf{\nabla_{x}}^2\Psi_{dm}&=&0,\\
\mathbf{\nabla_{x}}^{2} \phi(\mathbf{x},\tilde{\mathcal{D}})&=&4\pi G a^2\rho_{u}.
\nonumber
\end{eqnarray}
To obtain an exact solution we use the exact solutions to the three dimensional time dependent Schr\"{o}dinger equation by \cite{chand}. Exact solutions that \cite{chand} used are based on the group transformation method introduced by \cite{burgan79}. Based on this method, here we obtain the well-known solution of three-dimensional Schr\"{o}dinger equation, which is given by, 

\begin{eqnarray}
\label{dmpibox}
\psi^{\prime}(q,\tilde{\mathcal{D}})= \mathcal{A B C}e^{-iE\mathcal{D}}\sin(\frac{\pi}{d_{1}}n_{1}
q_{1})\sin(\frac{\pi}{d_{2}}n_{2} q_{2})\sin(\frac{\pi}{d_{3}}n_{3} q_{3}).
\end{eqnarray}
Here the initial location $\mathbf{q}$ is the Lagrangian coordinate of the matter element and moves
along the path $\mathbf{x}(\mathbf{q})$ and $\gamma(\mathbf{q})$ is the displacement
potential field. Also, $\mathcal{A B C}$ indicates the normalization constant for each spatial dimension while the values of $n_{i}$'s ($i=1$, $2$, $3$) are three integer numbers and they characterize the solution and, $E$ refers to the total energy. The mathematical description of the energy is given by,

\begin{eqnarray}
E= E_{1}+ E_{2}+ E_{3}=E_{0}\left(\frac{n^2_{1}}{d^{2}_{1}}+\frac{n^2_{2}}{d^{2}_{2}}+\frac{n^2_{3}}{d^{2}_{3}}\right),
\end{eqnarray}
in which $E_{0}\equiv\frac{\pi^2}{2}$. Now one may extend this into the general solution to,

\begin{eqnarray}
\Psi(x,y,z,\mathcal{D})&=& \mathcal{A B C}
e^{i\sqrt{\nu^{\prime}}\left(\frac{{\phi}_{v}}{\nu^{\prime}}-E_{0}\left(n^2_{1}+n^2_{2}+n^2_{3}\right)\mathcal{D}\right)}
\sin(\frac{\pi}{d_{1}}n_{1} \left(x+\mathcal{D}\nabla\gamma(q_{1})\right))
\sin(\frac{\pi}{d_{2}}n_{2}\left(y+\mathcal{D}\nabla\gamma(q_{2})\right))\nonumber\\
&&\sin(\frac{\pi}{d_{3}}n_{3} \left(z+\mathcal{D}\nabla\gamma(q_{3})\right)).
\label{darkmatterdensity}
\end{eqnarray}
In Fig. \ref{fig:darkmatter}, the evolution of the dark matter component is shown in terms of the linear growth
factor $\mathcal{D}$. As is seen from Fig. \ref{fig:darkmatter}, the scaled density $\chi_{dm}$ of the dark matter component increases with increasing growth factor $\mathcal{D}$. This indicates that dark matter density becomes prominent at the presentday $\mathcal{D}=1$ (or $z=0$).

\section{Example B: Baryon Dynamics in EdS Universe at $\delta\ll 1$ and $\lambda > \nu$}
\begin{figure}
\centering
\begin{tabular}{cc}
\includegraphics[width=0.4\textwidth]{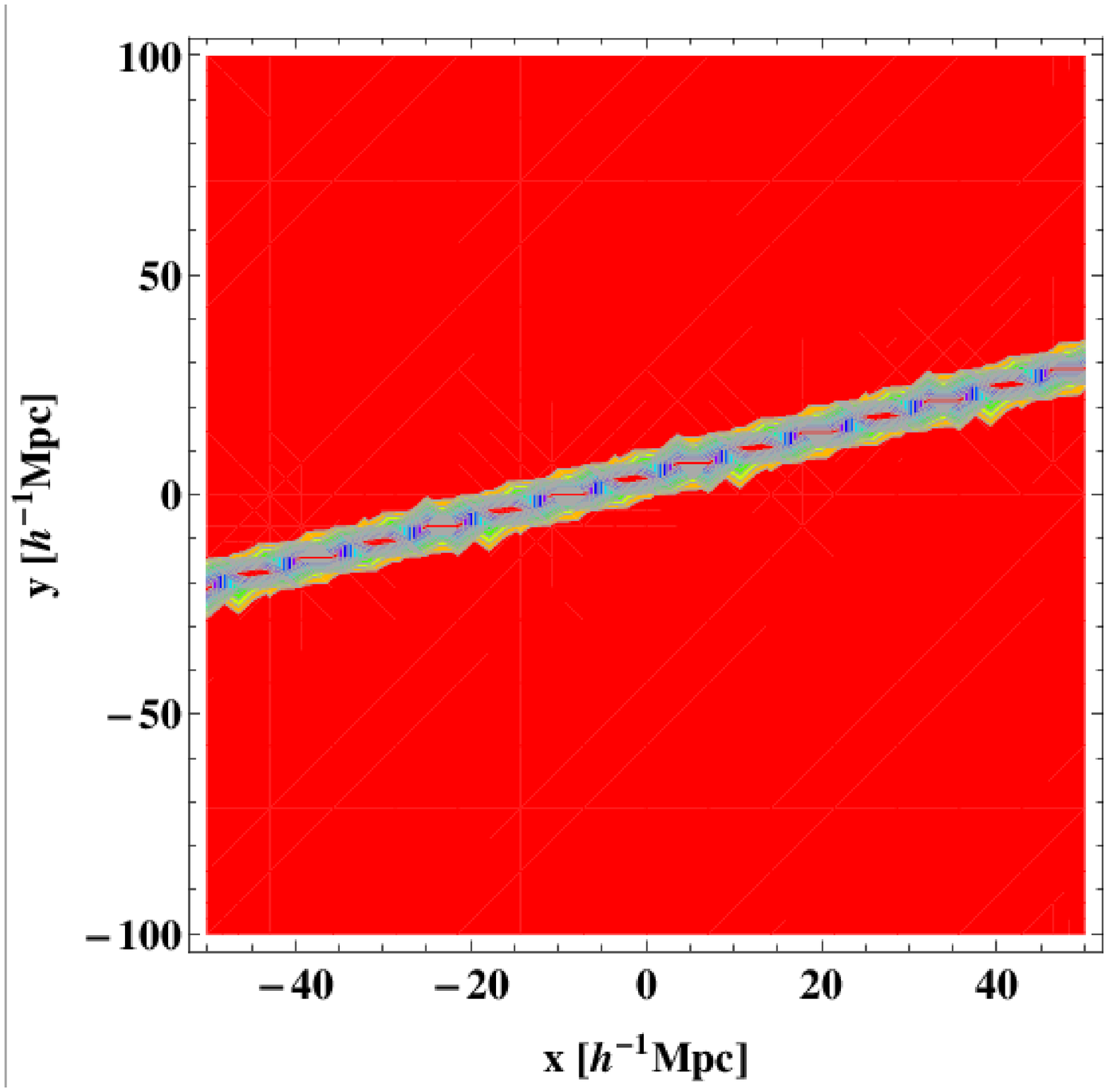}
\includegraphics[width=0.4\textwidth]{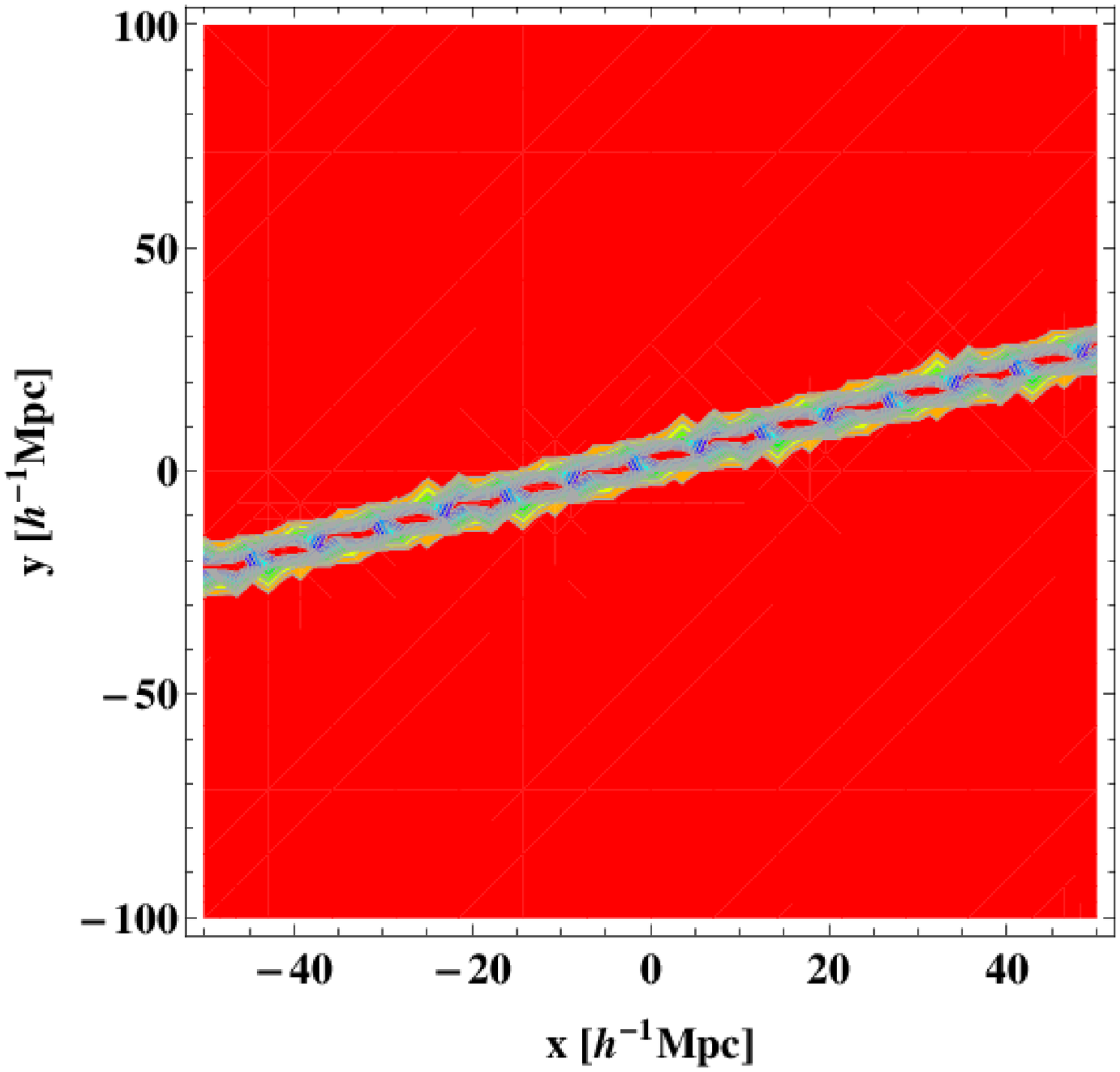}
\end{tabular}
\caption{Contour plots of the evolution of the scaled baryon density function
$\chi$ represented by the one-soliton solution of the reaction diffusion system (\ref{RD11bb}) from redshift
$z = 1$ to presentday redshift value $z=0$ (from left to right).}
\label{fig:onesola}
\end{figure}

\begin{figure}
\centering
\begin{tabular}{cc}
\includegraphics[width=0.4\textwidth]{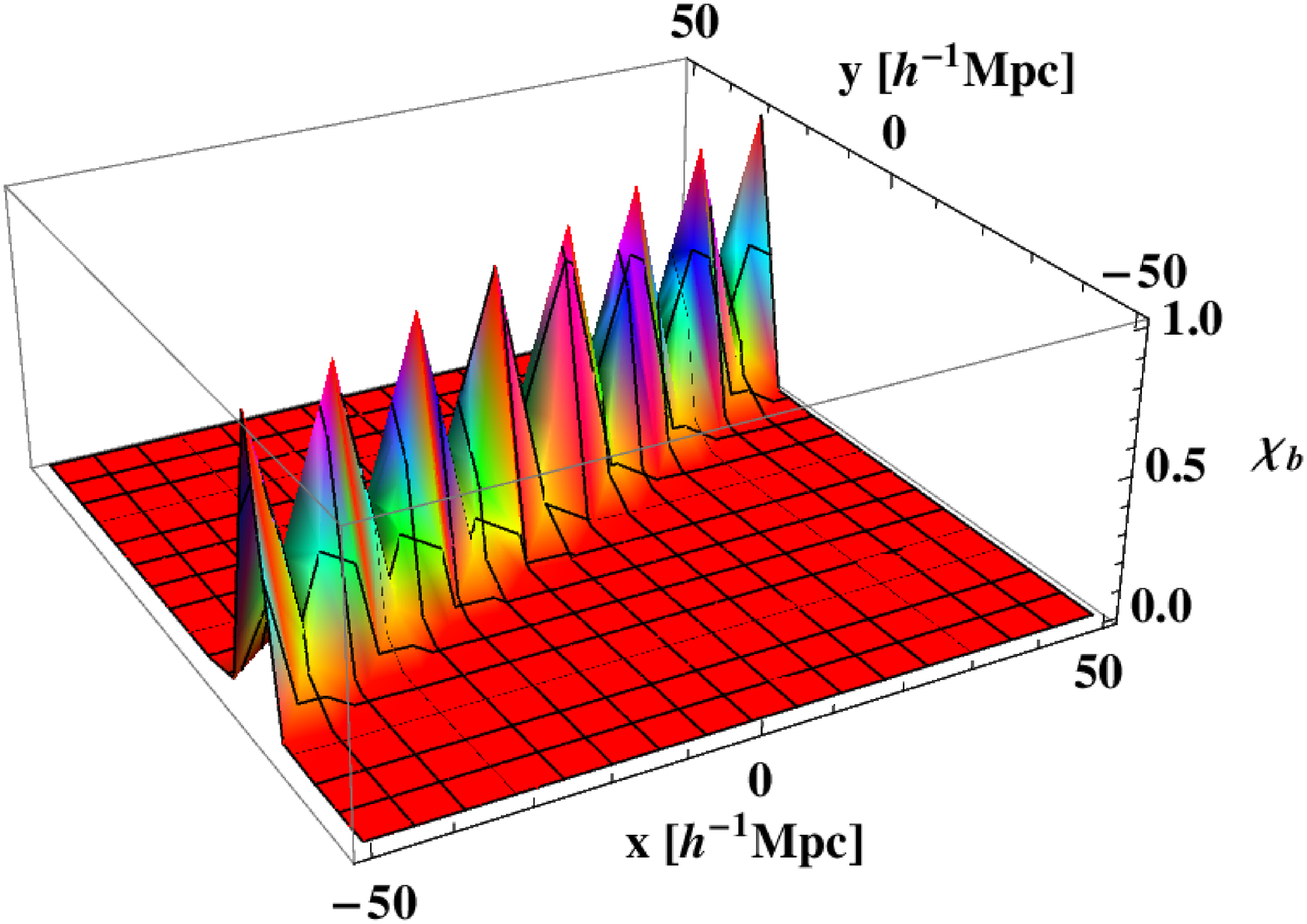}
\includegraphics[width=0.4\textwidth]{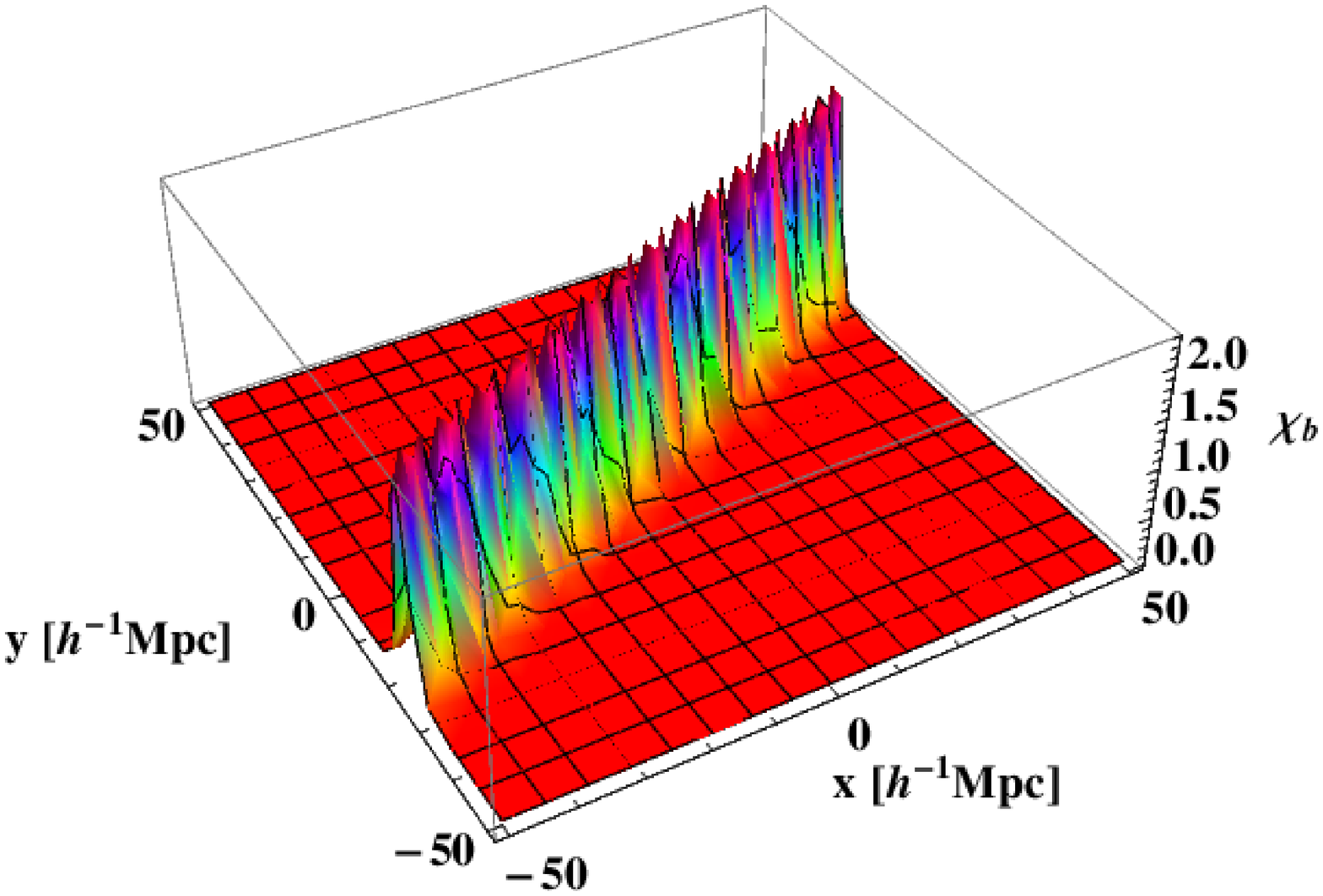}
\end{tabular}
\caption{Three-dimensional plots of the evolution of the scaled baryon density function
$\chi$ represented by the one-soliton solution of the reaction diffusion system (\ref{RD11bb}) from redshift
$z = 1$ to presentday redshift value $z=0$ (from left to right).}
\label{fig:onesolb}
\end{figure}

\begin{figure}
\begin{tabular}{cc}
\includegraphics[width=0.4\textwidth]{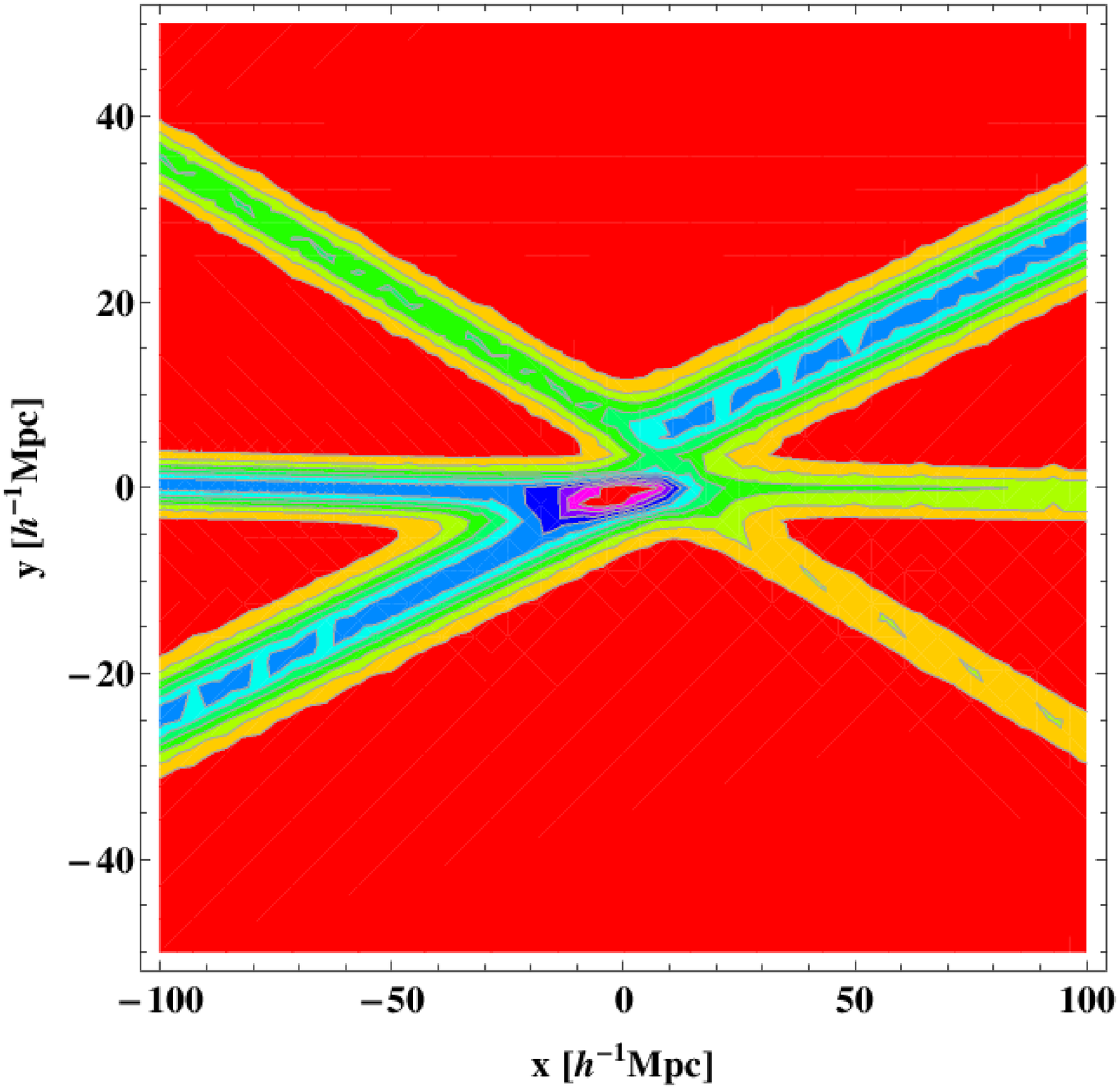}
\includegraphics[width=0.4\textwidth]{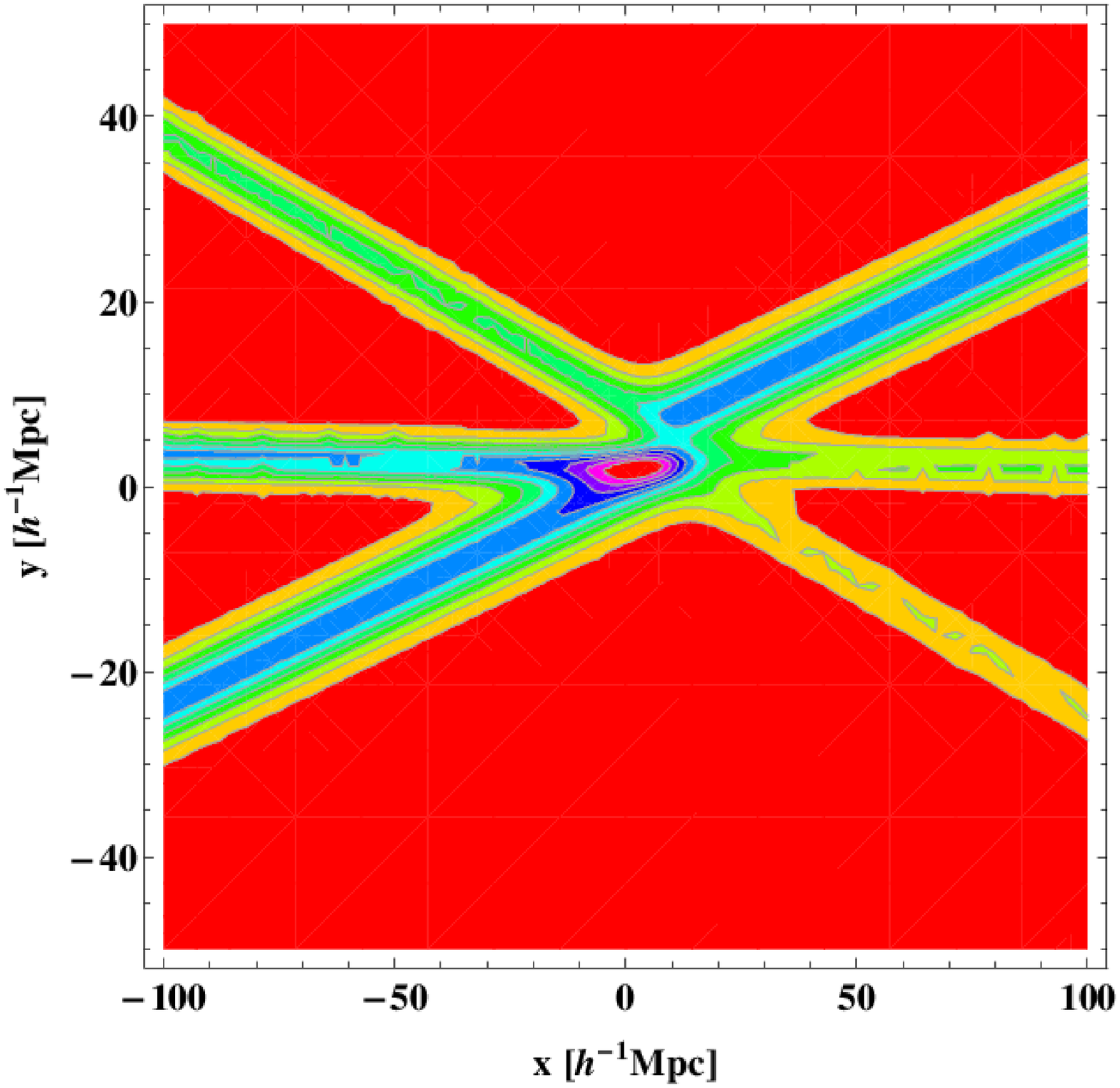}
\end{tabular}
\caption{Contour plots of the baryon density in terms of the two-soliton
solution based on system (\ref{RD11bb}) at redshift $z= 1$ and $z=0$ (from left to right).}
\label{fig:twosola}
\end{figure}

\begin{figure}
\begin{tabular}{cc}
\includegraphics[width=0.45\textwidth]{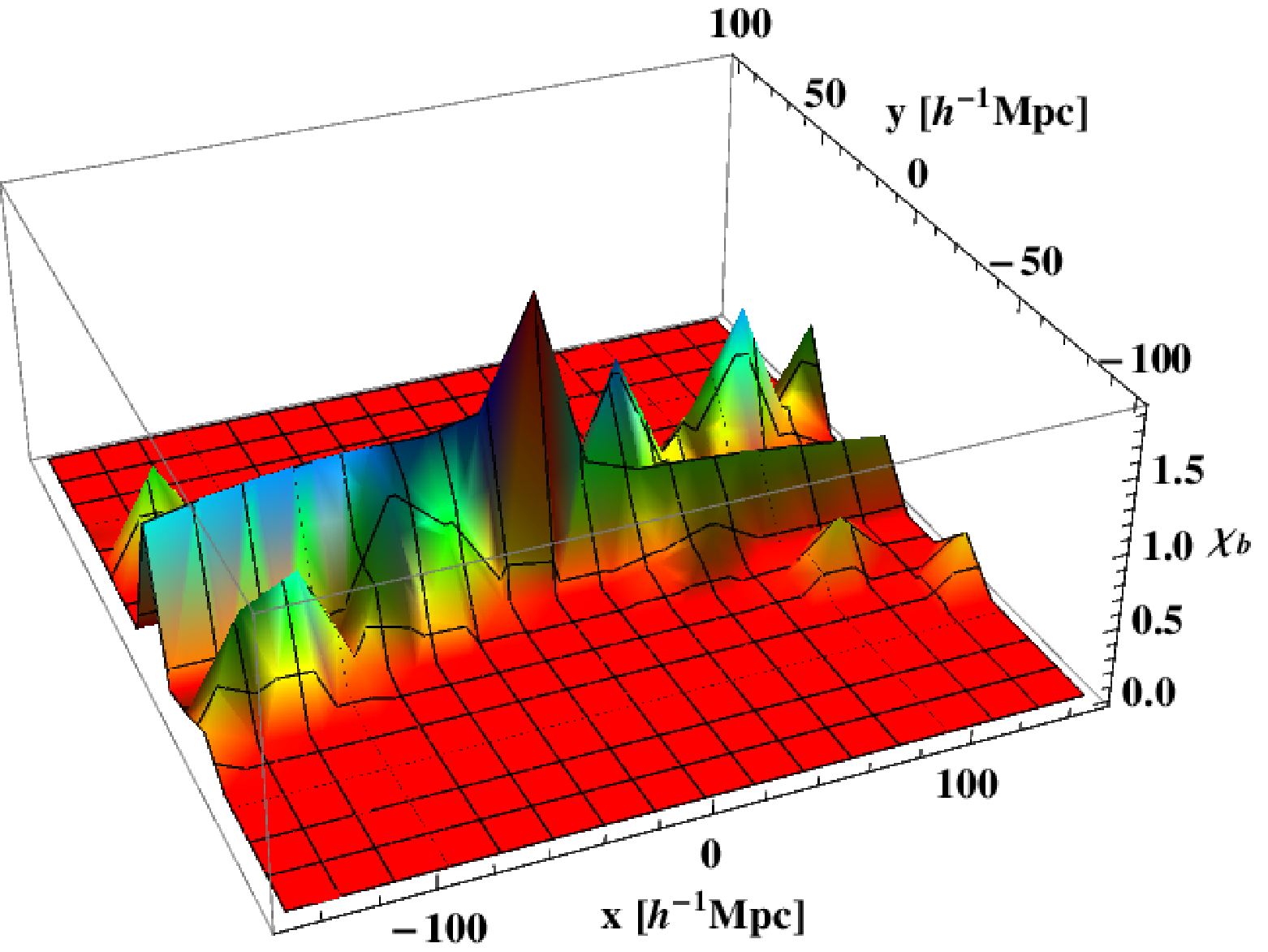}
\includegraphics[width=0.5\textwidth]{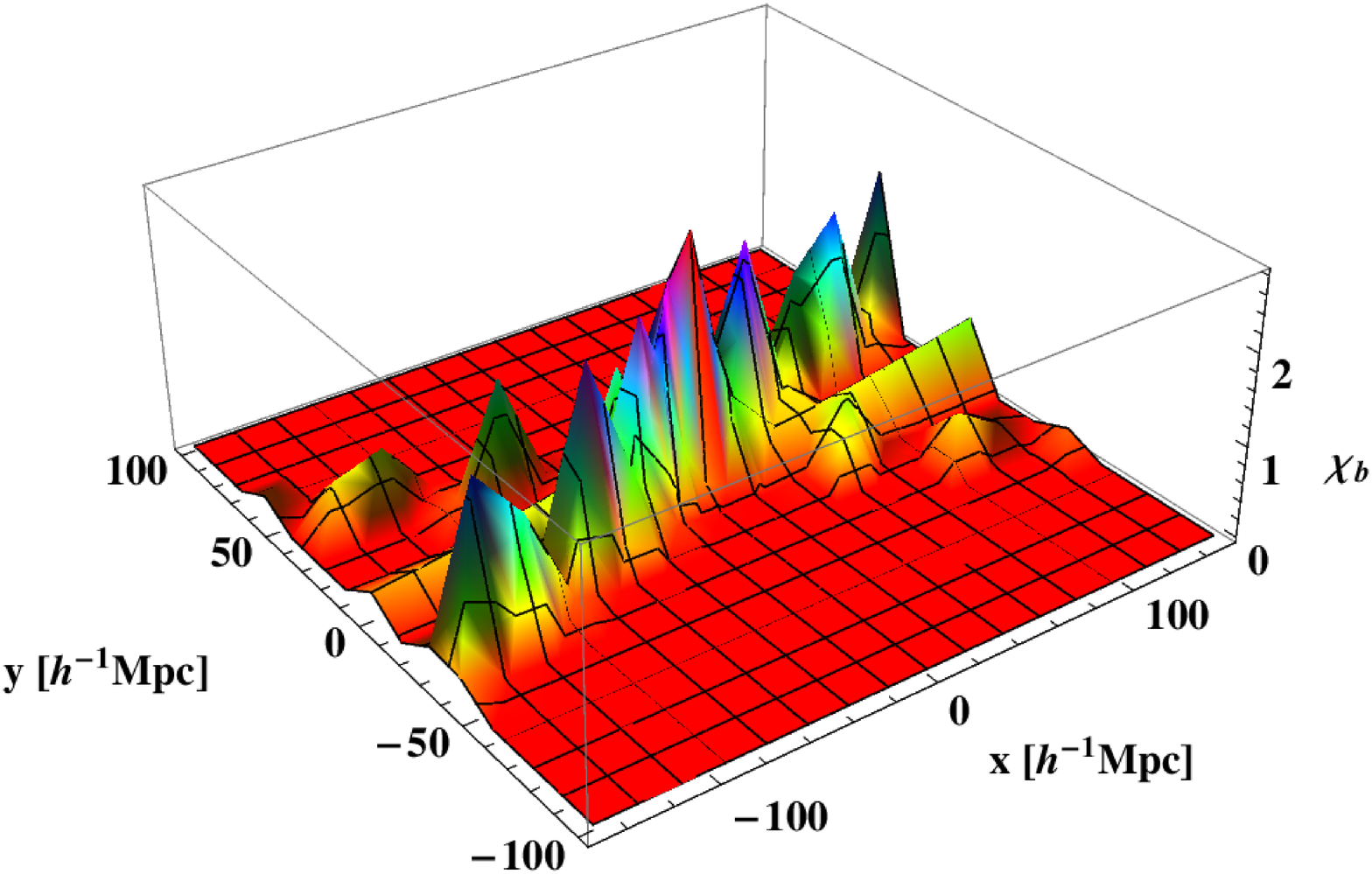}
\end{tabular}
\caption{Three-dimensional plots of the baryon density in terms of the two-soliton
solution based on system (\ref{RD11bb}) at redshift $z= 1$ and $z=0$ (from left to right).}
\label{fig:twosolb}
\end{figure}

Similar to its fluid dynamical counterpart, there are two possible wave mechanical descriptions of the baryon component. Here we provide a solution for the case of self-organization of the baryon component $\lambda > \nu$, and as such it can be modeled by a reaction diffusion system in the EdS universe,

\begin{subequations}
\label{reactionsysbcvefflack}
\begin{eqnarray}
\frac{\partial Q^{+}_{b}}{\partial \tilde{\tilde{\mathcal{D}}}}+\frac{1}{2}\nabla_{\mathbf{x}}^2Q^{+}_{b}
-{A^{2}(\tilde{\tilde{{\mathcal{D}}}})}\frac{\omega}{\nu^{\prime\prime}}{\ln|Q^{+}_{b}Q^{-}_{b}|}
Q^{+}_{b} &=&0,\\
-\frac{\partial Q^{-}_{b}}{\partial \tilde{\tilde{\mathcal{D}}}}+\frac{1}{2}\nabla_{\mathbf{x}}^2Q^{-}_{b}
-{A^{2}(\tilde{\tilde{{\mathcal{D}}}})}\frac{\omega}{\nu^{\prime\prime}}{\ln|Q^{+}_{b}Q^{-}_{b}|}Q^{-}_{b}&=&0,
\label{reactiondiffeqn2bex}\\
\mathbf{\nabla_{x}}^{2} \phi_{b}(\mathbf{x},\tilde{\tilde{\mathcal{D}}})&=& 4\pi G a^{2}\rho_{u}.
\label{reactiondiffeqn1bex}
\end{eqnarray}
\end{subequations}
It is a well known fact that in the linear regime and mildly linear regime the density perturbations become small
$\delta\ll 1$. Taking into account this fact, we can organize the logarithmic nonlinear dispersion term  $\ln|Q^{+}_{b}Q^{-}_{b}|$ in the reaction diffusion system (\ref{reactionsysbcvefflack}). Based on the linear regime with the density perturbations $\delta \ll 1$, the series expansion of the logarithmic function around the point $\chi_{b} \ll 1$ for the dispersion parameter in the system (\ref{reactionsysbcvefflack}) is obtained as,

\begin{eqnarray}
\ln |Q^{+}Q^{-}|&=&\ln \chi_{b}\nonumber\\&=& \left(\chi_{b}-1\right)-\frac{1}{2}\left(\chi_{b}-1\right)^{2}+\frac{1}{3}\left(\chi_{b}-1\right)^{3}-...
=\delta_{b}-\frac{\delta^{2}_{b}}{2}+\frac{\delta^{3}_{b}}{3}-\frac{\delta^{4}_{b}}{4}+\frac{\delta^{5}_{b}}{5}-...
\end{eqnarray}
\noindent
Here higher order terms can be omitted in the linear regime due to $\delta\ll 1$. Therefore, logarithmic nonlinearity is reduced to,

\begin{eqnarray}
\ln |Q^{+}Q^{-}|&=&\ln \chi_{b}\approx\delta_{b}.
\label{approxlog}
\end{eqnarray}
\noindent
Hence, the reaction diffusion system of the baryon component is reduced to a relatively simple form in the EdS universe,
\begin{subequations}
\label{RD11bb}
\begin{eqnarray}
\frac{\partial Q^{+}_{b}}{\partial {{\tilde{\tilde{{\mathcal{D}}}}}}}&+&\frac{1}{2}\nabla_{x}^2Q^{+}_{b}-A^{2}(\tilde{\tilde{\mathcal{D}}})\frac{\omega}{\nu^{\prime\prime}}\left(Q^{+}_{b}Q^{-}_{b}+1\right)Q^{+}_{b}=0,
\\
-\frac{\partial Q^{-}_{b}}{\partial {\mathcal{\mathcal{{{\mathcal{D}}}}}}}&+&\frac{1}{2}\nabla_{x}^2Q^{-}_{b}-A^{2}(\tilde{\tilde{\mathcal{D}}})\frac{\omega}{\nu^{\prime\prime}}\left(Q^{+}_{b}Q^{-}_{b}+1\right)Q^{-}_{b}=0,
\end{eqnarray}
\end{subequations}
\noindent
This system of equations can be solved by introducing the new functions as particular solutions,
\begin{eqnarray}
Q^{+}_{b}={\mathcal{Q}^{+}}_{b}e^{\frac{\omega}{\nu^{\prime
\prime}}\int^{\tilde{\tilde{\mathcal{D}}}}A^{2}(\eta)d\eta},\phantom{a}
Q^{-}_{b}={\mathcal{Q}^{-}}_{b}e^{-\frac{\omega}{\nu^{\prime\prime}}\int^{\tilde{\tilde{\mathcal{D}}}}A^{2}(\eta)d\eta}.
\label{RDgeneralsol}
\end{eqnarray}
Then system of equations (\ref{RD11bb}) is reduced the following form,

\begin{subequations}
\label{RD2}
\begin{eqnarray}
\frac{\partial \mathcal{Q}^{+}}{\partial {{\tilde{\tilde{{\mathcal{D}}}}}}}&+&\frac{1}{2}\nabla_{\mathbf{x}}^2\mathcal{Q}^{+}-A^{2}(\tilde{\tilde{\mathcal{D}}})\frac{\omega}{\nu^{\prime\prime}}
\mathcal{Q}^{+}\mathcal{Q}^{-}\mathcal{Q}^{+}=0,
\\
-\frac{\partial \mathcal{Q}^{-}}{\partial {{\tilde{\tilde{{\mathcal{D}}}}}}}&+&\frac{1}{2}\nabla_{\mathbf{x}}^2\mathcal{Q}^{-}-A^{2}(\tilde{\tilde{\mathcal{D}}})\frac{\omega}{\nu^{\prime\prime}}
\mathcal{Q}^{-}\mathcal{Q}^{+}\mathcal{Q}^{-}=0.
\end{eqnarray}
\end{subequations}
Our approach is to find the solution of (\ref{RD2}) by following the method developed
by \cite{Hirota71}. This solution leads us to the solution of equation (\ref{RD11bb}). When we apply the Hirota direct method to the reaction diffusion equation, the soliton solutions are obtained. The soliton solutions of the reaction diffusion system admit the exponentially growing and decaying components known as dissipatons. In this study to provide some illustrations, we then obtain one- and two- soliton solutions of the reaction diffusion system (\ref{RD11bb}) in order to show the evolution of baryon density $\chi_{b}$ in the case of self-organization. 
The one-soliton solution of the RD system has a relatively simple form,

\begin{eqnarray}
\chi_{b}=
\frac{\nu^{\prime\prime}}{A^2 \omega}\frac{k^2_{x}+k^2_{y}}{\cosh^2\left(k_{x} x+k_{y} y+\widetilde{\Omega} \mathcal{D}+\beta+\eta^{\pm}(0)\right)},
\label{rdonesol}
\end{eqnarray}
where $k_{x}$ and $k_{y}$ are represented as amplitude of the wave while $\zeta_{x}$ and $\zeta_{y}$ are velocities of the dissipative soliton,

\begin{eqnarray}
\widetilde{\Omega}&\equiv &\frac{\sqrt{\nu^{\prime\prime}}}{2}\left[k_{x}\zeta_{x}+k_{y}\zeta_{y}\right],\nonumber\\
\beta &\equiv& \frac{1}{2}\ln\left|\frac{A^2\omega}{4\nu^{\prime\prime}}
\frac{1}{k^{2}_{x}+k^{2}_{y}}\right|,\nonumber\\
k_{x}&\equiv &\frac{k^{+}_{1}+k^{-}_{1}}{2},\phantom{a} k_{y}\equiv \frac{m^{+}_{1}+m^{-}_{1}}{2},\nonumber\\
\zeta_{x}&\equiv &{k^{-}_{1}} - {k^{+}_{1}},\phantom{a}\zeta_{y}\equiv {m^{-}_{1}} - {m^{+}_{1}}.
\end{eqnarray}
The scaled density function shows the perfect soliton wave shape presented as filamentary type structure. When we change the parameters $k^{\pm}_{1}$ and $m^{\pm}_{1}$ we can see the different type of soliton wave types structures which bear a striking geometric resemblance to the filaments of the Cosmic Web. The one-soliton solution provides the information of the distribution of the scaled density of the baryon component in terms of the expanding scale factor or linear growth factor $a(t)=\mathcal{D}$, in other words, it shows the evolution of the scaled density function in the EdS Universe in Fig. ~\ref{fig:onesola} and Fig.~\ref{fig:onesolb}. Figs
\ref{fig:onesola} and \ref{fig:onesolb} demonstrate the one-soliton solution of the scaled baryon density in
which the density is plotted by choosing the adiabatic parameter as $\omega=5/3$ and the perturbation coefficients as $k^{+}_{1}=0.9$, ${k^{+}}_{1}=-0.5$, ${m^{+}}_{1}=0.3$, ${m^{+}}_{1}=-1.1$, ${\eta^{+}}(0)=-5$ and $\eta^{-}(0)=8$. In addition to the one-soliton solution, here we provide the two-soliton solution of the scaled baryon in
order to show how baryon component can show intricate structures with increasing order of soliton solutions in Fig. \ref{fig:twosola} and Fig. \ref{fig:twosolb} in the case of self-organization. In Fig. \ref{fig:twosola} and Fig. \ref{fig:twosolb}, the perturbation parameters are chosen as $\omega=5/3$,
$k^{+}_{1}=0.52$, $k^{-}_{1}=-0.2$, $k^{+}_{2}=0.4$, $k^{-}_{2}=-0.41$, $m^{+}_{1}=0.2$, $m^{-}_{1}=-0.01$,
$m^{+}_{2}=0.7$, $m^{-}_{2}=-1.6$, $\eta^{+}_{1}(0)=-6.5$, $\eta^{-}_{1}(0)=4$, $\eta^{+}_{2}=-6.5$ and
$\eta^{-}_{2}=4$. As is seen in Fig. \ref{fig:twosola} and Fig. \ref{fig:twosolb}, the scaled baryon density slightly increases with increasing redshift evolution. In the concept of large scale structure, this may indicate that the matter clumps merge through the bridges into the filaments. In the late time steps, these matter bridges become denser due to the merging of the matter into the high density regions/lumps in the linear regime.

\subsection{Conclusion}
In this study we introduce two methodologies in order to deal with the dynamical evolution of dark and baryon components of the Universe. This is achieved by obtaining the Schr\"{o}dinger type and reaction diffusion dynamical  forms by applying the two different Madelung transformations, based on the study of \cite{1964mhdp.book.....M}, to the cosmological fluid dynamical equations. Following this, here the quantum potential/pressure term is named as self-organization component of cosmic components based on the suggestions of \cite{bh93,bh94}. Note that since this self organization component  emerges out of the large scale cosmological fluid dynamical equations by using the Madelung transformation in the macroscopic scales, it is not the quantum potential. Then the Fisher information measure is introduced and compared to the nonlinear self-organization term od the systems via the lagrange constraint $\lambda$ of the Fisher measure and dispersion coefficient $\nu$. Following the studies discussing and deriving the dynamical importance of the Fisher information measure, here we show that Fisher information measures the dynamical behavior of the cosmological fluid by indicating self-organization or loss of self-organization depending on the comparison with the system's itself. Processing from this fact, the two different dynamical forms of the dark matter and baryon components are constructed in relatively simple frameworks in the linear regime due to the vanishing effective potential. As a result, in the case of loss of self-organization $\lambda < \nu$, the dark matter wave equation is reduced to the free particle Schr\"{o}dinger equation in the linear regime, while the baryon matter wave equation shows the special kind of nonlinear differential equation called the log-law nonlinear Schr\"{o}dinger equation. On the other hand, when the Fisher information measure increases $\lambda > \nu$, the dark matter presents the heat system of wave equations while the baryon component obeys the coupled reaction diffusion system with log-law nonlinearity. Here, to provide some examples we present analytical solutions of dark and baryon matter components in different dynamical characteristics given by Fisher information cases. The dark matter Schr\"{o}dinger type wave form is solved by using the particle in a box method and it is shown that there are some similarities between the Zel'dovich formalism and this dark matter wave form. 
To solve the coupled reaction diffusion system with log-law nonlinearity, we use a special methodology called the Hirota direct method. Due to the nature of the Hirota method, we obtain perturbative solutions of the baryon component. 
When we increase the order of the perturbations in the Hirota method from one-soliton solution to $N$-soliton solutions, these waves show striking similarity to the intricate structure of filamentary type features of the Cosmic Web in $2+1$ dimensions in the EdS Universe in the linear regime.
\bibliographystyle{mn2e}
\bibliography{bib}
\bsp
\label{lastpage}
\end{document}